\documentclass[prl,twocolumn,superscriptaddress]{revtex4}
\usepackage{graphicx}
\usepackage{graphics}      
\usepackage{bm}    
\usepackage{amsmath}     
\usepackage{amsfonts}
\usepackage{amssymb}
\usepackage{latexsym}  
\begin{document}
\title{Boltzmann-Shannon Entropy:Generalization and Application}
\newcounter{count}
\author{C.G Chakrabarti}
\email{cgc_math@rediffmail.com}\affiliation{Department of Applied
Mathematics,Calcutta University,Kolkata 700009,India}
\author{I.Chakrabarty}
\email{indranilc@indiainfo.com} \affiliation{Heritage Institute of
Technology, Kolkata,700107, India}\affiliation{Bengal Engineering
and Science University, Howrah, W.B, India}
\date{\today}
\begin{abstract}
The paper deals with the generalization of both Boltzmann entropy
and distribution in the light of most-probable interpretation of
statistical equilibrium. The statistical analysis of the
generalized entropy and distribution leads to some new interesting
results of significant physical importance.
\end{abstract}
\maketitle
\section{\bf Introduction}
Boltzmann was first to provide the statistical definition of
entropy linking the concept of entropy with molecular disorder or
chaos $^1$. Boltzmann entropy is the key to the foundation of
statistical mechanics and is, in fact, the basis of all
statistical concepts of entropy. The concept of probability which
is vital for a statistical theory,however, has not come out clear
with Boltzmann entropy. For, the thermodynamic probability or
statistical weight appearing in Boltzmann entropy, is not a
probability, it is an integer. The statistical equilibrium as
defined by Boltzmann and Planck to be the most probable state
achieved by maximizing the thermodynamic probability brought with
it certain opaqueness $^2$.The object of the present paper is to
modify Boltzmann entropy in order to introduce the notion of
probability distribution in Boltzmann statistics. In this
objective we have first considered a classical system and have
reduced the Boltzmann entropy to the form of Shannon entropy $^3$,
not in terms of probabilities, but in terms of occupation numbers
of different energy states of the system. This form of entropy is
called Boltzmann-Shannon entropy has been modified in the light of
most probable interpretation of statistical equilibrium. The
modified entropy called Boltzmann-Shannon cross entropy has led
two important results. The first is the probability distribution
of the macro states consistent with Einstein's inversion of
Boltzmann principle $^3$. The second is the equivalence of
information and negentropy consistent with the Bernoulli's
negentropy principle of information $^4$. The most probable
interpretation of statistical equilibrium has led to a generalized
form of Boltzmann distribution involving prior probabilities. The
appearance of prior probabilities makes the results interesting
for both physical and non physical systems.
\section{\bf Boltzmann-Shannon Entropy and Probability }
Boltzmann entropy of a system is defined by,
\begin{eqnarray}
S=k\ln W
\end{eqnarray}
where k is the Boltzmann constant and W, called the thermodynamic
probability or statistical weight, is the total number of
microscopic states or complexions compatible with the macroscopic
state of the system. The thermodynamic probability W appearing in
Boltzmann entropy (1) is not a probability, it is an integer. We
may however, ask for the probability $P(A_n)$ of any macroscopic
state $A_n$ (say).This probability may be represented as the
fraction representing the ratio of the statistical weight $W_n$ to
the sum of statistical weights of all the macroscopic states that
are compatible with the given constraints:
\begin{eqnarray}
P(A_n)=\frac{W_n}{ \sum_{A_{n}} W_n}
\end{eqnarray}
Since for large $W_n$ $^2$
\begin{eqnarray}
(W_n)_{max}=W_{total}=\sum_{A_n}W_n
\end{eqnarray}
the probability (2) may be replaced by
\begin{eqnarray}
P(A_n)=\frac{W_n}{(W_n)_{max}}=\exp{\frac{S-S_{max}}{k}}
\end{eqnarray}
where $S_{max}$ is the maximum value of the entropy S.It was this
form of Boltzmann principle that was used successfully by Einstein
$^5$ in his study of thermodynamic fluctuations and its various
applications. In this way Einstein introduced the probability
distribution by inverting Boltzmann principle. Note that in this
approach entropy comes first and probability comes later on.\\
In the present paper we shall follow a different path. We shall
first introduce the probability distribution of macrostates and
find the expression of entropy consistent with the general
mathematical theory of entropy $^6$. Before we do that we consider
a classical system and reduce the Boltzmann entropy to the form of
entropy, not in terms of probabilities but in terms of occupation
numbers of the different energy states of the system. Let the
system under consideration consists of $N$ molecules classified
into n energy states $E_i$ $(i=1,2,...,n)$ with corresponding
occupation numbers $N_i(i=1,2,...n)$. The system is assumed to be
isolated system characterized by fixed values of the total energy
and number of molecules:
\begin{eqnarray}
\sum_{i=1}^{n} N_{i}= N (fixed)\nonumber\\
\sum_{i=1}^{n} N_{i}E_{i}= E (fixed)
\end{eqnarray}
The macroscopic state of the system is given by the set of
occupation numbers $A_n=[N_1,N_2,....,N_n]$. Thus the statistical
weight of the macroscopic state $A_n=[N_1,N_2,....,N_n]$ is given
by
\begin{eqnarray}
W(A_n)=\frac{N!}{\prod_{i=1}^n N_{i}!}
\end{eqnarray}
representing the total number of microscopic states of the system.
For large $N_i(i=1,2...n)$, using Stirling's approximation,
Boltzmann entropy S with statistical weight (6) is reduced to the
form
\begin{eqnarray}
S=k\ln \frac{N!}{\prod_{i=1}^n N_{i}!}\thickapprox -kN\sum_{i=1}^n
p_{i}\ln p_{i}
\end{eqnarray}
where $p_i=\frac{N_i}{N}$ is the relative frequency and for large
$N$, it is the probability that a molecule lies in the ith energy
state $E_i$. The expression
\begin{eqnarray}
H(p_1,p_2,....,p_n)=-k\sum_{i=1}^n p_{i}\ln p_{i}
\end{eqnarray}
appearing in the right hand side of (7) is the Shannon entropy
measuring the uncertainty associated with the probability
distribution $(p_1,P_2,...,p_n)$ $^3$. Thus for large classical
system Boltzmann entropy is proportional to the Shannon entropy
and as such the Shannon entropy defined by (8) is also a measure
of molecular disorder of the system. In terms of occupation
numbers the expression (7) can be written as
\begin{eqnarray}
S=-k\sum_{i=1}^n N_{i}\ln N_{i}+k\ln N
\end{eqnarray}
The second term in the right hand side of (9) is a constant for
constant number of molecules constituting the system. So for
variational purpose this constant may be dropped and we can write
Boltzmann entropy for classical system in the form
\begin{eqnarray}
S=-k\sum_{i=1}^n N_{i}\ln N_{i}
\end{eqnarray}
Note that (10)has the same functional form of Shannon entropy (8)
and is defined over the non probabilistic distribution
$[N_1,N_2,....N_n]$. Due to this similarity with Shannon entropy
(8) we call it Boltzmann-Shannon entropy. In the next section we
are going to generalize the Boltzmann-Shannon entropy (10) along
with its physical or thermodynamic significance.\\
\section{\bf Probability-Distribution of Macrostates: Boltzmann-Shannon Cross-Entropy}
According to Boltzmann and Planck the thermodynamic equilibrium is
defined as the most-probable state. The thermodynamic probability
$W(N_1,N_2,....,N_n)$ of the macrostate $A_n=[N_1,N_2,...,N_n]$ is
not a probability, it is an integer. So the thermodynamic
equilibrium obtained by the maximization of thermodynamic
probability $W(N_1,N_2,....,N_n)$ or equivalently Boltzmann
entropy (1) may lead to some confusions $^2$. To find out the
most-probable state we have to determine first the probability
distribution of the macrostate $A_n=[N_1,N_2,...,N_n]$ on the
basis of the prior information or data. The occupation numbers
$[N_1,N_2,...,N_n]$ are assumed to be a set of random variables in
view of the many-body aspect of the system. Let
$P(A_n)=P[N_1,N_2,...,N_n]$ be the probability distribution of
$[N_1,N_2,...,N_n]$. Let the mean or averages of occupation
numbers $[N_1,N_2,...,N_n]$ be known:
\begin{eqnarray}
\sum _{R_{N,n}} N_i P(N_1,N_2,...,N_n)= \bar{N_i}
\end{eqnarray}
where $i=1,2...n$ and $R_{N,n}$ is the set of non negative
integers satisfying the condition,
\begin{eqnarray}
N_1+N_2+....+N_n=N
\end{eqnarray}
The mean value $\bar{N_i}(i=1,2...n)$ given by (11) constitute
constraints about the system. Note that the thermodynamic
probability or statistical weight $W(N_1,N_2,...,N_n)$  given by
(6) is the prior information about the macrostate
$A_n=[N_1,N_2,...,N_n]$ of the system .An appropriate measure of
uncertainty or entropy about the system is given by Bayesian
entropy $^7$.
\begin{eqnarray}
S_B=\sum_{R_{N,n}}P(N_1,N_2,.....,N_n)\ln\frac{P(N_1,N_2,.....,N_n)}{W(N_1,N_2,...,N_n)}
\end{eqnarray}
Our problem is to estimate the probability distribution
$P(N_1,N_2,.....,N_n)$ under the prior information (6) and the
constraints (11). We can do this by the generalization of Jaynes'
Maximum-entropy principle $^8$. According to Jaynes $^8$ the best
estimate of the probability distribution $P(N_1,N_2,.....,N_n)$
corresponds to the maximization of the Bayesian entropy (13)
subject to the constraints (11) and the normalization condition:
\begin{eqnarray}
\sum_{R_{N,n}}P(N_1,N_2,.....,N_n)=1
\end{eqnarray}
The best estimate of the probability, $P(N_1,N_2,.....,N_n)$ is
then given by the multinomial distribution $^{7,9}$
\begin{eqnarray}
P(N_1,N_2,.....,N_n)= \frac{N!}{\prod_iN_i!}(p_i^0)^{N_i}
\end{eqnarray}
where $p_i^0$ $(i=1,2,...n)$ is the prior probability that a
molecule lies in the i-th energy state and it is determined from
the prior information or constraint (11). In the existing
literature the multinomial distribution of the macrostate
$A_n=[N_1,N_2,...,N_n]$ is usually assumed without any physical
justification. We have, however, provided an information-theoretic
method based on generalized maximum-entropy principle which takes
account of the prior available constraints and information about
the system.\\
Assuming $N_i (i=1,2,...,n)$ to be very large and using Stirling's
approximation we can reduce the logarithm of the probability
$P(N_1,N_2,.....,N_n)$ to the form
\begin{eqnarray}
k\log P(N_1,N_2,.....,N_n)=-Nk\sum_{i=1}^n p_i\ln\frac{p_i}{p_i^0}
\end{eqnarray}
where $p_i= \frac{N_i}{N}$, $(i=1,2,...,n)$ and for the large N,
it is the probability that a molecule lies in i-th energy state
$E_i$. Note that (16) is a generalization of (3) and is in fact,
the measure of relative entropy. Again except the multiplication
constant $(-Nk)$ the expression (16) is known as Kullback-Leibler
relative information or simply Kullback cross-entropy giving a
measure of directed divergence between the probability
distributions $[p_1,,p_2,...,p_n]$ and $[p_1^0,,p_2^0,...,p_n^0]$
$^{10}$.\\
Let us now transform (16) in terms of the occupation numbers
$[N_1,N_2,...N_n]$. The priori probabilities $p_i^0$ are the
parameters of the multinomial distribution (15). As we have stated
$p_i^0$ are to be determined in terms of the available constraints
(11) that is, in terms of $\bar{N_i}$ $(i=1,2,...,n)$. An unbiased
estimate of $p_i^0$ is given by $p_i^0= \frac{\bar{N_i}}{N}$. Then
replacing $p_i$ by $\frac{N_i}{N}$ and $p_i^0$ by
$\frac{\bar{N_i}}{N}$, we have
\begin{eqnarray}
k\ln P(N_1,N_2,.....,N_n)=-k\sum_{i=1}^n
N_i\ln\frac{N_i}{\bar{N_i}}
\end{eqnarray}
which is a generalization of Boltzmann-Shannon entropy (10). It
is, in fact, a measure of relative entropy defined over the
non-negative integers. Since $P(N_1,N_2,..,N_n)\leq 1$ the
quantity (17) is, however , negative. We shall call the negative
of (17) that is , the quantity
\begin{eqnarray}
-k\ln P(N_1,N_2,.....,N_n)=k\sum_{i=1}^n
N_i\ln\frac{N_i}{\bar{N_i}}
\end{eqnarray}
as the generalized Boltzmann entropy. The left hand side of (18)is
the probabilistic entropy of the macrostate
$A_n=[N_1,N_2,...,N_n]$. The right hand side has the same form as
that of Kullback cross-entropy $^{10}$ and due to this similarity
we shall call this expression Boltzmann-Shannon cross-entropy
defined over the set of non-negative integers
$[N_1,N_2,...,N_n]$.In the following we are
going to study its physical or thermodynamical significance.\\
Let us assume that the averages $\bar{N_i}$ $(i=1,2,..,n)$
correspond to the thermodynamic equilibrium values of $N_i$
$(i=1,2,..n)$, then it is easy to show that $^{11}$
\begin{eqnarray}
-k\sum_{i=1}^n N_i\ln \frac{N_i}{\bar{N_i}}= S-S_{equil}
\end{eqnarray}
where S is the entropy of the system at non equilibrium state
$[N_1,N_2,...,N_n]$ and $S_{equil}$ be that of equilibrium state
$[\bar{N_1},\bar{N_2},...,\bar{N_n}]$. From (18) and (19) we have,
\begin{eqnarray}
-k\ln P(N_1,N_2,...,N_n)=S_{equil}-S
\end{eqnarray}
The left hand side of (20) which we have stated to represent the
probabilistic entropy of the macrostate $[N_1,N_2,...,N_n]$ is
also the measure of information obtained about macrostate
$[N_1,N_2,...,N_n]$ after its realization $^6$. The right hand
side of (20) is the negentropy of the system at the state
$[N_1,N_2,...,N_n]$. The relation (20) thus implies the
equivalence of information and negentropy and is consistent with
the Brillouin's negentropy principle of information $^4$. The
relation (20) also provides another important result. From (20) we
can write the probability of macrostate $[N_1,N_2,...,N_n]$ as
\begin{eqnarray}
P[N_1,N_2,...,N_n]=\exp [\frac{S-S_{equil}}{k}]
\end{eqnarray}
consistent with Einstein's result (4) obtained by inverting
Boltzmann principle.
\section{\bf Generalized Boltzmann Distribution and Applications}
In Boltzmann statistics the distribution law of thermodynamic
equilibrium is determined by maximizing the thermodynamic
probability $W(N_1,N_2,...N_n)$ given by (6) or equivalently the
Boltzmann-Shannon entropy (10) subject to the constraints (5).The
entropy (10) subject to the constraints (15). The maximization
yields the Boltzmann distribution
\begin{eqnarray}
p_i= \frac{e^{-\beta E_i}}{Z(\beta)}
\end{eqnarray}
where
\begin{eqnarray}
Z(\beta)=\sum_{i=1}^n e^{-\beta E_i}
\end{eqnarray}
and the parameter $\beta$ may be identified with the inverse
temperature by the relation $\beta=\frac{1}{kT}$, T being the
absolute temperature of the system. According to most probable
interpretation of the statistical equilibrium the maximum of the
probability $P(N_1,N_2,..,N_n)$ or equivalently $\ln
P(N_1,N_2,..,N_n)$ given by (16) subject to the constraints
corresponds to the statistical equilibrium. We have then the
generalized Boltzmann distribution,
\begin{eqnarray}
p_i=p_i^0 [\frac{e^{-\beta E_i}}{Z(\beta)}]
\end{eqnarray}
where
\begin{eqnarray}
Z(\beta)=\sum_{i=1}^n e^{-\beta E_i}
\end{eqnarray}
The difference with the usual or old Boltzmann distribution (22)
comes out from the multiplicative factor $p_i^0$, the prior
probability. The appearance of prior probabilities makes the
problem complex and it is difficult to determine the prior
probabilities in statistical mechanics or in any other branch of
science $^{12}$ $^{13}$. If no state is more preferable to other,
it is then reasonable to assume that all the prior probabilities
$p_i^0$  are equal to one another so that $p_i^0=\frac{1}{n}$,
$(i=1,2,..n)$. This is Laplace's principle of insufficient
knowledge $^{13}$ $^{14}$. According to Jaynes' $^8$ this is the
state of maximum prior ignorance. In the case of equal prior
probabilities the most -probable distribution reduces to the form
\begin{eqnarray}
p_i=\frac{1}{n}\frac{e^{-\beta E_i}}{Z(\beta)}
\end{eqnarray}
which except the multiplicative constant $\frac{1}{n}$ is the
Boltzmann-distribution derived earlier. For most-probable state of
thermodynamic equilibrium with equal priori probabilities
$\frac{1}{n}$ the entropy (7) becomes
\begin{eqnarray}
S_{equil}= -kN \sum_{i=1}^n p_i\ln p_i\nonumber\\
=kN [\ln n + (\beta E+\ln Z(\beta))]
\end{eqnarray}
On the other hand, with unequal prior probabilities, the entropy
of thermodynamic equilibrium corresponding to the probabilities
distribution (24) is given by
\begin{eqnarray}
S_{equil}^0 = -Nk\sum_{i=1}^n p_i^0\ln p_i^0 +Nk(\beta E+\ln
Z(\beta))
\end{eqnarray}
Since,$\ln  n\geq-\sum_{i=1}^np_i^0\ln p_i^0$, we have
\begin{eqnarray}
S_{equil}\geq S_{equil}^0
\end{eqnarray}
implying that the thermodynamic equilibrium with unequal prior
probabilities $p_i^0$ does not correspond to the maximum entropy
or maximum disorder of the system. This is a violation of the
existing physical law and is due to unequal prior probabilities
$^{15}$.\\ We now consider a physical problem where unequal prior
probabilities appear and make things different from those with
equal prior probabilities. Let us consider a collection of N
linear harmonic oscillators all with frequency $\upsilon$. The
energy levels of a linear harmonic oscillator is given by,
\begin{eqnarray}
E_i = (i-\frac{1}{2})h\upsilon
\end{eqnarray}
where $(i=1,2,..)$ and h is Planck's constant. When the prior
probabilities are all equal then the total energy of the system is
is given by
\begin{eqnarray}
E=\frac{h\upsilon}{2}+\frac{h\upsilon}{e^{\beta h\upsilon}-1}
\end{eqnarray}
 Now if the collection is of two dimensional
harmonic oscillators, the situation becomes different. In this
case prior probabilities $p_i^0$ increases linearly with i, so
that we can write $^{16}$
\begin{eqnarray}
p_i^0=\frac{i}{C}
\end{eqnarray}
where $(i=1,2,..)$ and $C$ is a necessary to make $p_i^0$
probability in true sense. In the case of two dimensional
oscillators $^{16}$
\begin{eqnarray}
E_i=ih\upsilon
\end{eqnarray}
where (i=1,2,..) and a bit of calculation gives the total energy
of the system as $^{16}$
\begin{eqnarray}
E=h\upsilon+\frac{2h\upsilon}{e^{\beta h\upsilon}-1}
\end{eqnarray}
where the zero-energy level has been changed from that of linear
harmonic oscillators.
\section{\bf Conclusion:}
In the present paper we have made an attempt to generalize both
Boltzmann entropy and Boltzmann distribution in the light of
Boltzmann Planck most-probable interpretation of statistical
equilibrium. We have obtained some interesting new results
different from the old ones and tried to find out their physical
significance. Let us state some of the main results along with
their merits.\\
(i) The present method of determination of the probability
distribution of macrostates is more direct and transparent than
the old method of inverting Boltzmann principle.\\
(ii) Boltzmann-Shannon cross-entropy obtained as a generalization
of Boltzmann-Shannon entropy is defined over the set of
non-negative integers $[N_1,N_2,..,N_n]$. The relation (18),
however shows that it has probabilistic meaning and is consistent
with the probabilistic foundation of entropy and
information.\\
(iii) The most-probable interpretation of statistical equilibrium
leading to the generalized Boltzmann distribution (24) involves
prior probabilities $p_i^0$. The appearance of prior probabilities
makes the result interesting different from the existing ones,
sometimes in violation of the existing physical laws, for example,
the maximum entropy or disorder for statistical equilibrium
$^{15}$.\\
(iv) The importance of prior probability in a physical system
namely a collection of two dimensional harmonic oscillators has
been investigated. Prior probabilities also play a significant
role in statistical mechanical modelling of ecosystems
$^{17}$.\\
Boltzmann-Shannon entropy is a classical one, its generalization,
however, leads to some new results of important physical
significance. Finally, the mathematical simplicity of the paper
which is independent of any mechanical or statistical models and
postulates $^{18}$ is an advantageous point of the theory.
\section{\bf Acknowledgement}
The authors wish to thank the learned referee whose comments helps
the better exposition of the paper.
\section{\bf References}
$[1]$ L.Boltzmann, Lectures on Gas Theory (English
Translation),(University of California, Berkley, 1964)\\
$[2]$ B.H. Lavenda, Statistical Physics: A probabilistic Approach,
(Wiley and Sons, New-York,1991).\\
$[3]$ C.F. Shannon and W.Weaver: Mathematical Theory of
Communication, University of Illinois, Urbana, 1949.\\
$[4]$ L.Brillouin, Science and Information Theory, (Academic
Press,
New-York,1956).\\
$[5]$ A. Einstein: Zur allegemeinen malckulaven Theorie de Warme
.Ann. De Phys 14,354(1904).\\
$[6]$ J.Aczel and Z.Doroc'zy: On Measures of Information and their
Characterization. (Academic Press, New-York,1975).\\
$[7]$ J.N. Kapur, Maximum-Entropy Models in Science and
Engineering (Wiley Eastern, New-Delhi, 1990).\\
$[8]$ E.T.Jaynes, Phys. Rev. 106, 620 (1957).\\
$[9]$ C.G. Chakrabarti,and M.De: Ind.J. Phys, 76A,573 (2002)\\
$[10]$ S.Kullback: Information Theory and Statistics, ( Wiley and
Sons, New-York, 1959).\\
$[11]$ C.G.Chakrabarti and K.De: J.Biol. Phys. 23,(1997), 169.\\
$[12]$ K.G.Denbigh and J.S. Denbigh : Entropy in relation in
relation to incomplete knowledge, (Cambridge Univ.Press,
Cambridge,1985)\\
$[13]$ A.Katz : Principles of Statistical Mechanics, (W.H Freeman
and Co. San-Francisco, 1967).\\
$[14]$ R.Baierlein: Atoms and Information Theory, ( W.H. Freeman
and Co. San-Francisco, 1967).\\
$[15]$ P.T. Landsberg: J.Stat.Phys. 35,159(1984)\\
$[16]$ J.R.Lindsay: Introduction to Physical Statistics, (Wiley
and
Sons, New-York,1941).\\
$[17]$ P.N.Augar: Dynamics and Thermodynamics of Hierarchically
Organized Systems, (Pergamon Press, Oxford,1989)\\
$[18]$ R.Penrose: Foundation of Statistical Mechanics, (Pergamon
Press, Oxford,1970).

\end{document}